\documentclass[]{IEEEtran}
\usepackage[dvips]{color} 

\usepackage[dvips]{graphicx}
\ifCLASSINFOpdf
\usepackage[pdftex]{graphicx}
\else
\fi
\begin{document}
%
\title{Dipolar field effect on microwave oscillation in a domain wall spin-valve}
%
%
%
\author{Katsuyoshi~Matsushita\thanks{Katsuyoshi Matsushita, Email: k-matsushita@aist.go.jp}, 
Jun~Sato and
Hiroshi~Imamura\\
Nanotechnology Research Institute (NRI),\\
Advanced Industrial Science and Technology (AIST),\\
AIST Tsukuba Central 2, Tsukuba, Ibaraki 305-8568, Japan.}

%
%

\markboth{Journal of \LaTeX\ Class Files,~Vol.~6, No.~1, January~2007}%
{Shell \MakeLowercase{\textit{et al.}}: Bare Demo of IEEEtran.cls for Journals}
%



\maketitle

\begin{abstract}
We examined dipolar field effects on the microwave generation  
in the domain wall spin-valve by solving simultaneously the Landau-Lifshitz-Gilbert and Zhang-Levy-Fert diffusion equations.
By numerically analyzing dipolar field dependence, we showed that the microwave generation needs the dipole-dipole interaction for a 180$^\circ$ domain wall and the amplitude of the microwave voltage signal depends strongly on the exchange length. 
In order to design a microwave generator using the domain wall spin-valve with high efficiency, we propose that the materials with short exchange length are preferred. 
\end{abstract}


%
\IEEEpeerreviewmaketitle

%
%
%
%
\IEEEPARstart{I}{n} recent years, much effort has been devoted for development 
of a microwave generator using a nano-scale spin-transfer torque oscillator\cite{Katine:2000,Tsoi:2000,Kiselev:2003,Rippard:2004,Covington:2004,Krivorotov:2004,Kasa:2005,Mancoff:2005}.
As a powerful candidate of the nano-scale spin-transfer oscillator,
a domain wall spin-valve is proposed\cite{Doi:2007}. 
The domain wall spin-valve consists of a magnetic nano-multilayer, the main part of which is 
a nano-oxide-layer (NOL) sandwiched by ferromagnetic layers\cite{Fuke:2008}.
The NOL contains a lot of fine contacts filled up with ferromagnetic metal. The contacts enable dc-current to be conducted by bias voltage.   
Under the dc-current, a microwave voltage signal is switched on when domain walls in the contacts are constructed.
This implies that the microwave voltage signal is considered to originate from dynamics of the domain walls driven by the dc-current in the contact. 

\begin{figure}[tb]
    \begin{center}
     \includegraphics[scale=0.5]{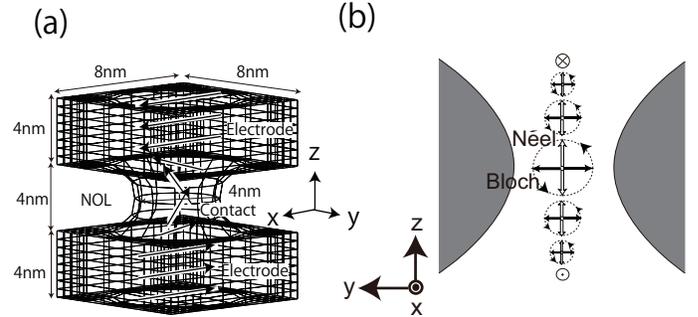}
    \end{center}
    \caption{The schematic picture of the model of a geometrically confined domain wall (a) and the rotation of the magnetic structure (b). White and black arrows denote the magnetic structures for the N\'{e}el and Bloch walls, respectively. The dotted circle denote the rotation of the local magnetization configuration.}
    \label{fig:model}
   \end{figure}

  The model of the domain wall spin valve is shown in Fig.~\ref{fig:model}(a).
Based on the model we previously reproduces the microwave generation in the domain wall spin-valve under dc-current by a micromagnetic simulation technique\cite{Matsushita:2009}.
In the simulation the spin-transfer torque drives a uniform rotation of local magnetizations as shown in Fig.~\ref{fig:model}(b). In the rotation the magnetic structure oscillates between the N\'{e}el and Bloch walls. 
The oscillation generates the microwave voltage signal by inducing an oscillation of spin accumulation which contributes voltage drop. 
However the reason why the oscillation of the magnetic structure induces the microwave voltage signal is still unclear because the voltage drop induced by the spin accumulation mainly depends not on the magnetic structures but on the thickness of the domain wall.
The mechanism should be clarified
in order to design a microwave generator using the domain wall spin valve with high efficiency.

In the present work, we propose a scenario of a breathing mode excitation in the domain wall, which relates the rotation of the magnetic structure with the oscillation of the spin accumulation and is induced by both the dipole-dipole interaction and spin-transfer-torque. The scenario is as follows: 
The spin-transfer torque induces the rotation of the magnetic structure \cite{Matsushita:2009,Tatara:2004,Ono:2008}.
Then a breathing mode, which is an oscillation of the thickness of the domain wall, appears 
because the dipole-dipole interaction effectively introduces a difference between the thickness of the N\'{e}el and Bloch walls. 
As a result, the dc-current is converted to an ac voltage signal because the resistance of the domain wall due to the spin accumulation is proportional to squared inverse of its thickness\cite{Simanek:2001}.  

The scenario predicts strong dependence of magnitude of the microwave voltage signal on the exchange length, $l_{\rm ex}$, defined by $\sqrt{J_{\rm dd}/2K_{\rm d}}$, where $J_{\rm dd}$ and $K_{\rm d}$ are coupling constants of the exchange and dipole-dipole interaction between local magnetizations, respectively, because the dipolar field yields the breathing mode.  
In the present paper, to confirm the prediction of our scenario, we simulated dynamics of the local magnetization configuration and spin accumulation in the domain wall spin valve by solving simultaneously the Landau-Lifshitz-Gilbert\cite{Hubert:1998} and Zhang-Levy-Fert diffusion equations\cite{Zhang:2002} and investigated the dipolar field effects by artificially controlling the exchange length.
We considered the values of the exchange length of CoFe, Permalloy and infinity.
Based on numerically analyzing exchange length dependence, we showed that the efficient microwave generation needs the strong dipole-dipole interaction for the 180$^\circ$ domain wall geometrically confined in the contact\cite{Bruno:1999} and the magnitude of the microwave voltage signal depends strongly on the exchange length. Our results suggests that the materials with short exchange length is preferred in order to develop the microwave generator using the domain wall spin-valve with high efficiency. 

   The system consists of two magnetic electrode layers and a
   non-magnetic insulator layer sandwiched by the electrodes as shown in Fig.~\ref{fig:model}(a).  The
   insulator layer contains magnetic contact which geometrically confines
   a domain wall.  The system is divided into about 1500 hexahedral
   finite elements, where spin accumulation and demagnetization field
   are evaluated.  The size of magnetic metal electrode at the bottom
   and top layers is 8 $\times$ 8 $\times$ 4 nm.  The shape of the
   contact between two electrodes is a rotated elliptic arch around the
   center axis perpendicular to the layers, the diameters of the bottom
   and center layers of which are set at 6 and 4 nm, respectively.  
   The size of the system is larger than the diffusion length of 2nm of CoFe\cite{Moyerman:2005} at room temperatures. Thus the electronic system is diffusive at least. 
   We deal with the electron system in diffusive limit, unlike a previous ballistic treatment\cite{ohe:2006}.
  
   The local magnetizations are expressed by the classical spins on
   the simple cubic lattice with the lattice constant of $a= 0.4$nm.  
   The Hamiltonian ${\cal H}$ is given by
   \begin{eqnarray}
    &&{\cal H} = -J_{\rm dd} \sum_{\left<i,j\right>} \vec S_{i} \cdot \vec
    S_{j} + J_{\rm sd} \sum_i \vec S_{i} \cdot \delta \vec m_i 
    \nonumber \\ 
    \mbox{}&&\hspace{0em}
    + \frac{K_d}{4\pi} \sum_{i} \vec S_{i} \cdot \int d\vec r 
    \left\{ \frac{\hat 1}{|\vec r_{i}|^3}-3\frac{\vec r_{i} \otimes \vec r_{i}}{|\vec r_{i}|^5} \right\} \cdot \vec S(\vec r),
    \label{Hamiltonian}
   \end{eqnarray}
   where $\vec{r}_{i}$ represents the relative coordinate of the $i$-th site from the position $\vec r$; $\vec
   S_{i}$, the classical Heisenberg spin with absolute value of unity and
   $\delta \vec m_{i}$, local spin accumulation
   density at the $i$-th site.  
   The first term in the right hand side of Eq.~(\ref{Hamiltonian})
   expresses the exchange interactions between local magnetizations at
   nearest neighbor sites.  The exchange coupling
   constant denoted by $J_{\rm dd}$
   is related to the exchange stiffness constant in continuous limit,
   $A$, and a lattice constant, $a$, with $J_{\rm dd} = 2aA$.  We fix
   $J_{\rm dd}$ at 0.04 eV which is of the order of the transition
   temperature of the typical material, $T_c \sim$ 0.8(Permalloy)-1.2(CoFe) $\times 10^3 $K,
   and the material dependence of $J_{\rm dd}$ is neglected because the dependence is much weaker than other material parameters and because the spin-transfer torque depends only on $J_{\rm dd}$ through the small correction of the order of the small parameter defined by $1/J_{\rm dd}\tau_{\rm sf}$\cite{Zhang:2004}, where $\tau_{\rm sf}$ is a spin relaxation time by a spin-orbit coupling. 
   
   The second term expresses the $s$-$d$ exchange interactions between local magnetizations and the spin accumulation at each site\cite{Zhang:2004}.
   The $s$-$d$ exchange coupling constant $J_{\rm sd}$ is set at 0.1 eV in accordance with Ref.\cite{Zhang:2002}. The term reproduces the spin-transfer torque induced by the spin accumulation. In the present paper, we adopted the Zhang-Levy-Fert diffusion equations\cite{Zhang:2002} for the evaluation of the spin accumulation. 
  
   The third term denotes the dipole-dipole interaction energy.  
   In order to deal with the dipole-dipole interaction in a complex
   system shape, we adopt a finite element - boundary element (FEM-BEM)
   hybrid method~\cite{Fredkin:1990} on the spin field on the continuum space,
   $\vec S(\vec r)$, which is defined for each position, $\vec r$, by the
   spin field extrapolated from the lattice sites to the position.  The
   exchange length for the dipole-dipole interaction is $l_{\rm ex}
   =\sqrt{J_{\rm dd}/2K_d}$ nm. We examined the cases of the exchange length of
   3nm(CoFe), 5nm(Permalloy) and $\infty$ by controlling artificially $K_{\rm d}$ in order to investigate dipolar field effects.

   In this paper we assume the system size of 4nm is larger than the mean free path.
   In this case, the local spin accumulation density.
   $\delta \vec m_{i}$, is determined by solving the 
   following  Zhang-Levy-Fert diffusion equation\cite{Zhang:2002},
   \begin{eqnarray}
    &\frac{\partial}{\partial t} \delta \vec m(\vec r) =
    \nabla \left\{ \beta \vec S(\vec r) \vec j_{\rm e}(\vec r)  
    +\hat A(\vec S(\vec r)) \delta \vec m(\vec r)  \right\} 
    \label{eq:ZLF1}\nonumber\\
    &\ \hspace{1.3cm}
    + \frac{J_{\rm sd}}{\hbar}\delta \vec m(\vec r)
    \times \vec S(\vec r) + \frac{\delta \vec m(\vec r)}{\tau},\\
    &\hat A(\vec S(\vec r)) = 2D_0\left[\hat 1 - \beta^2\vec S(\vec r)
    \otimes \vec S(\vec r)\nabla\right], 
    \label{eq:ZLF2} 
   \end{eqnarray}
   where $\vec j_e$, $C_0$, $D_0$, $\beta$ and $\tau$ denote electronic current, 
   conductivity, diffusion constant, polarization of resistivity and 
   relaxation time due to a spin-orbit interaction respectively. 
   Equations~(\ref{eq:ZLF1})-(\ref{eq:ZLF2}) are solved numerically
   with combining continuous equation for electronic current, 
   $\nabla \vec j_e = 0 \label{eq:continous}$. 
   $C_0$ and $\beta$ are taken to be those for conventional ferromagnets
   as $C_0 =$70$\Omega$nm and $\beta=0.65$, respectively. $D_0$ and
   $\tau$ are obtained, respectively, by the Einstein relation, $C_0 =
   2e^2N_{\rm F}D_0$ and $\lambda = \sqrt{2\tau D_0(1-\beta^2)}$ for given spin diffusion
   length $\lambda$, density of states at the Fermi level, $N_{\rm F}$
   and electron charge $e$. Here we employ $\lambda$=12nm and $N_{\rm
   F}=$7.5nm$^{-3}$eV$^{-1}$.
   
   We adopted an adiabatic approximation where the relaxation time of the spin accumulation is much shorter than that of the local magnetization. In that case, we can neglect dynamics of the local magnetization in the determination of the spin accumulation and consider the steady solution. 
   For the boundary condition of the spin accumulation, we artificially adopt
   $\delta \vec m = \vec 0$ at the top and bottom layers,
   ignoring any parasitic resistance.
   On the other boundaries, the natural boundary condition is employed.
   In the case,
   the simulated spin accumulation distribution is nonuniform and concentrates 
   on the contact region.
   We adopt current density of 0.01 mA/nm$^2$ where the top and bottom surface.
   The direction of current is aligned in the $z$-direction.
   
   Substituting the solution of Eqs.~(\ref{eq:ZLF1}) and (\ref{eq:ZLF2})
   into Eq.~(\ref{Hamiltonian}), we evaluate the Hamiltonian of
   Eq.~(\ref{Hamiltonian}).  The effective magnetic field for
   $\vec{S}_{i}$ is given by $\partial {\cal H}/\partial \vec S_i$ and 
   the dynamics of $\vec{S}_{i}$ is determined
   by the following Landau-Lifshitz-Gilbert equation,
   \begin{equation}
    \frac{d}{dt} \vec S_i 
     = \frac{\gamma}{1+{\alpha}^2} \vec S_i \times
     \left(  \frac{\partial {\cal H}}{\partial \vec S_i} + \alpha \vec
      S_i \times \frac{\partial {\cal H}}{\partial \vec S_i}\right),
   \end{equation}
   were $\gamma$ and $\alpha$ denote the gyromagnetic ratio and the
   Gilbert damping constant.  The equation is numerically solved by a
   quaternion method\cite{Visscher:2002}.  The time step $\Delta t$ and
   $\alpha$ are set at 3.4 $\times$ 10$^{-2}$ fs and 0.02,
   respectively. The number of sites, $N_{\rm s}$ is about $10^4$.  
   For simplicity, we use antiparallel boundary condition of the
   spins in order to simulate the experimental situation with a
   180$^\circ$ domain wall.  On the boundary, $\vec S$'s are fixed at
   (-1,0,0) on the top electrode and (+1,0,0) on the bottom electrode
   except for the boundary surface between the magnetic and insulator
   layers. The directions of spins, $x$, $y$ and $z$ are defined as shown
   in Fig.~\ref{fig:model}(a).
   
\begin{figure}[tb]
    \begin{center}
     \includegraphics[scale=0.9]{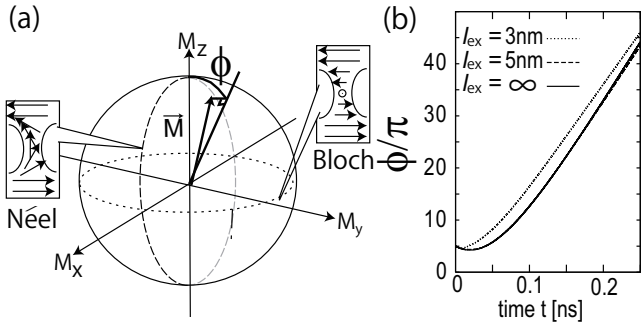}
    \end{center}
    \caption{The relation between $\vec M$ and magnetic structures (a) and the time evolution of $\phi$ for $l_{\rm ex}=$ $3$nm(CoFe), $5$nm(Permalloy), $\infty$. The dashed and doted lines in the $M_x$-$M_z$ and $M_x$-$M_y$ planes denote the order parameters for the N\'{e}el and Bloch walls where $M_y$ and $M_z$ is zero, respectively. (The definition of $\vec M$ is exactly given by Eq.~\ref{eq:magnetization}).}
    \label{fig: phi}
   \end{figure}

   To analyze the dynamics of the domain wall, especially 
   the rotation of magnetic structures, as shown in Fig.~\ref{fig: phi}(a) we calculate a vector, $\vec M$ which is the ferromagnetic order parameter scaled with its absolute value, 
   \begin{eqnarray}
    \vec M = \frac{ \vec M_{\rm op}}{ |\vec M_{\rm op}|}, \label{eq:magnetization}\\
    \vec M_{\rm op} = \frac{1}{N}\sum_i \vec S_i,
   \end{eqnarray}
   where $N$ is the number of sites of local magnetizations.
   We show $\phi={\rm arctan}(M_y/M_z)$ as a function of time $t$ in Fig.~\ref{fig: phi}(b).
   In the present setting the $\phi$ almost linearly increases by spin transfer torque after enough time is spent.
  The angular velocities for all cases finally reach values of 1 $\times 10^2$ Ghz.
   Because as shown in Fig.~\ref{fig: phi}(a) the domain wall for $\phi=n\pi$ and for $\phi=\pi(1/2+n)$, respectively, correspond to the N\'{e}el and Bloch walls, where $n$ is any integer, the linear increasing of $\phi$ suggests a magnetic structure rotation of the domain wall including the N\'{e}el and Bloch walls as discussed above. 
   
   \begin{figure}[tb]
    \begin{center}
     \includegraphics[scale=0.5]{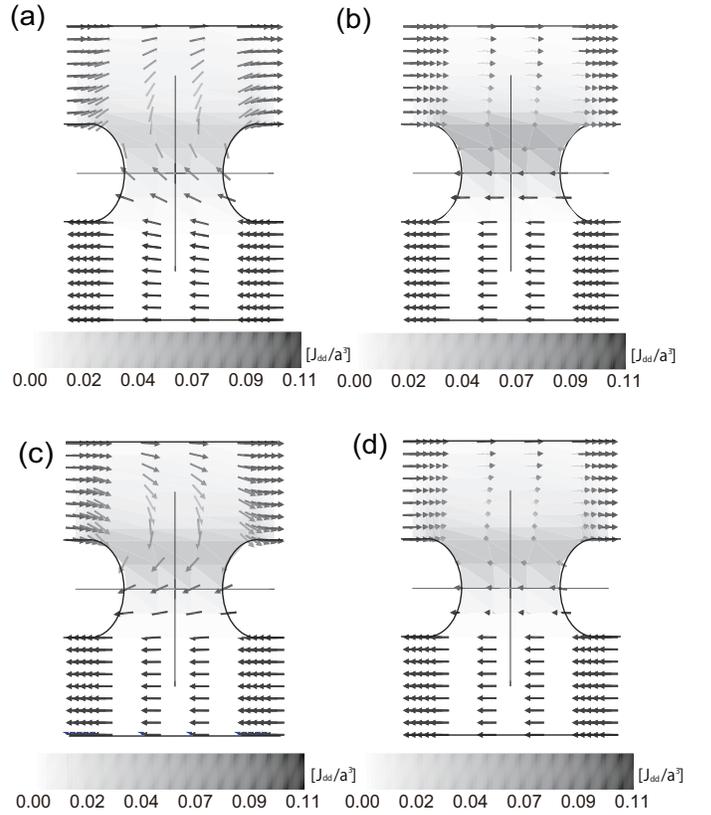}
    \end{center}
    \caption{The local magnetization configuration snapshots in the $x$-$z$ plane including the center of the system and exchange energy density for (a) $l_{\rm ex}= 3$nm with a N\'{e}el wall, (b) $l_{\rm ex}= 3$nm with a Bloch wall, (c) $l_{\rm ex}= \infty$ with a N\'{e}el wall and (d) $l_{\rm ex}= \infty$ with a Bloch wall.}
    \label{fig:energy density}
   \end{figure}
   Figures~\ref{fig:energy density}(a)-(d) show snapshots of exchange energy density of the local magnetization for $l_{\rm ex}$= 3nm and $\infty$ with the N\'{e}el and Bloch walls. As shown in Figs.~\ref{fig:energy density}(a) and (b) for $l_{\rm ex}=$3nm the exchange energy density depends on whether the domain wall is the N\'{e}el and Bloch ones. In fact the maximum values of the exchange energy density of the N\'{e}el and Bloch are 0.08 and 0.11 [$J/a^3$], respectively. The difference of the exchange energy density in the magnetic structures of the domain wall suggests that the resistance depends on $\phi$ and the resistance oscillation is induced by the rotational motion of $\phi$ because both of the exchange energy and resistance due to potential drop of the spin accumulation is roughly proportional to squared inverse of the thickness of the domain wall\cite{Simanek:2001}.
 On the other hand as shown in Figs.~\ref{fig:energy density}(c) and (d) for $l_{\rm ex}=\infty$ the exchange energy density almost does not depend on a magnetic structure of the walls. In fact the maximum value of the exchange energy density is the almost same value of about 0.1[$J/a^3$] for both the N\'{e}el and Bloch walls.  In this case, the resistance can not oscillates during the rotational motion of $\phi$. 

These results mean that the rotational motion driven by the spin-transfer torque induces a breathing mode which is an effective oscillation of the thickness of the domain wall. 
The breathing mode excitation induces the resistance oscillation of the system. 
As a result the microwave oscillation is expected because the oscillation of the resistance under a dc current corresponds an oscillation of voltage between the both ends of the contact. 
The breathing mode is excited only for the finite $l_{\rm ex}$ because the effective thickness dependence on the magnetic structure of the domain wall is induced by the dipolar field. 
Thus we speculate that the magnitude of the voltage signal becomes small as increasing $l_{\rm ex}$ and reaches zero at $l_{\rm ex}=\infty$.
   
   Let us to directly examine the oscillation of the resistance.  
   The electric field due to the spin accumulation is given by 
   \begin{equation}
    \vec E(\vec r) = \frac{1}{2C_0}
     \left(\vec j_{\rm e}(\vec r)+
      2D_0\left[\hat 1 + \beta \vec \sigma \cdot \vec S(\vec r)\right]
      \delta \vec m(\vec r)\right).
     \label{eq:voltage}
   \end{equation}
   \begin{figure}[tb]
    \begin{center}
     \includegraphics[scale=1.2]{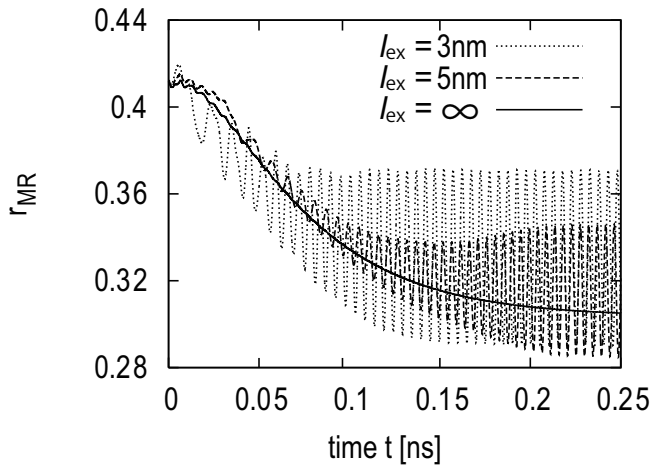}
    \end{center}
    \caption{The time evolution of $r_{\rm MR}$ for $l_{\rm ex}=$ $3$nm(CoFe), $5$nm(Permalloy), $\infty$.}
    \label{fig:MR}
   \end{figure}
From the electric field we evaluate time dependence of the resistance. Figure.~\ref{fig:MR} shows the resistance ratio $r_{\rm MR} = R(t)/R_0-1$ as a function of the time for $l_{\rm ex}= 3$, 5, $\infty$nm, here $R(t)$ and $R_0$ denote resistance and resistance without a domain wall. As speculated above the oscillation of $r_{\rm MR}$ is observed for finite $l_{\rm ex}$'s and the oscillation of $r_{\rm MR}$ is absent for infinite $l_{\rm ex}$. The magnitude of the oscillation of the resistance increases with decreasing the exchange length. From the discussion so far we conclude that the enhancement of the signal for short exchange length is because a strong dipolar field in a case of a short exchange length induces the large difference of the thickness of the domain wall between the N\'{e}el and Bloch walls. These results suggest that the magnitude of the microwave voltage signal is enhanced by adopting the material with short $l_{\rm ex}$ if one design a domain wall spin-valve.
   
Finally, we summarize our conclusions.
We examined the dipolar field effect on the microwave generation  
in the spin-valve by solving simultaneously the Landau-Lifshitz-Gilbert and Zhang-Levy-Fert diffusion equations.
Our results suggests that the dipole-dipole interaction crucially affects the microwave generation.
By numerically analyzing the exchange length dependence, we showed that the efficient microwave generation needs the strong dipole-dipole interaction and the amplitude of the microwave voltage signal depends strongly on the exchange length. We also showed that the microwave voltage signal is induced by the breathing mode originating from the dipolar field. In order to develop the microwave device with the domain wall spin-valve with high efficiency, we suggest from the results that the materials with short exchange length are preferred so that the breathing mode with large amplitude is excited. 

The authors thank M.~Doi, H.~Iwasaki, M.~Ichimura, K.~Miyake, H.~Ohotori, M.~Takagishi, K.~Seki
  M.~Sahashi, M.~Sasaki, T.~Taniguchi and N.~Yokoshi for useful
  discussions.  The work was supported by NEDO and MEXT.Kakenhi(19740243).
\ifCLASSOPTIONcaptionsoff
  \newpage
\fi

\bibliographystyle{IEEEtran}

%








\end{document}